\documentclass{PoS}

\def\0{\over } \def\1{\vec } \def\2{{1\over2}} \def\4{{1\over4}}
\def\5{\bar }
\def\6{\partial }
\newcommand{\bea}{\begin{eqnarray}}
\newcommand{\eea}{\end{eqnarray}}
\newcommand{\be}{\begin{equation}}
\newcommand{\ee}{\end{equation}}

\def\g{g_{\rm eff}}
\def\geff{g_{\rm eff}}

\def\tr{\,{\rm tr}\,}

\newcommand{\muMS}{\bar\mu_{\rm MS}}

\newcommand{\im}{{\rm Im}\,}

\title{Thermodynamics of QCD at large quark chemical potential }

\ShortTitle{Thermodynamics of QCD at large quark chemical potential }

\author{Andreas Gerhold\\
Department of Physics, North Carolina State University, Raleigh, NC 27695, USA\\
E-mail: \email{agerhol@unity.ncsu.edu}}

\author{Andreas Ipp\\
ECT*, Villa Tambosi, I-38050 Villazzano Trento, Italy\\
E-mail: \email{ipp@ect.it}}

\author{\speaker{Anton Rebhan}\\ 
  Institut f\"ur Theoretische Physik, Technische Universit\"at Wien, A-1040 Vienna, Austria\\
        E-mail: \email{rebhana@hep.itp.tuwien.ac.at}}

\abstract{We review the existing weak-coupling results on the thermodynamic
potential of deconfined QCD at small and large quark chemical potential
and compare with results from lattice gauge theory as well
as the exactly solvable case of large-$N_f$ QCD.
We also discuss the new analytical results on non-Fermi-liquid effects
in entropy and specific heat as well as in dispersion laws
of quark quasiparticles
at large quark chemical potential.}

\FullConference{29th Johns Hopkins Workshop on Current Problems in Particle Theory: Strong Matter in the Heavens\\
		 1-3 August 2005\\
		 Budapest}

\begin{document}

\section{Introduction}

Deep in the deconfined phase of QCD, with either temperature $T$ or
quark chemical potential $\mu$ much larger than the QCD scale
$\Lambda_{\rm QCD}$, asymptotic freedom eventually leads to a
strong coupling constant $g$ that is sufficiently small to
permit the use of perturbation theory for calculating
the thermodynamics of quarks and gluons under extreme
conditions. Indeed, much effort has been invested in calculating
the thermodynamic potential of QCD at high temperature. It
is by now known up to and including order $g^6 \log(g)$
\cite{Arnold:1995eb,Zhai:1995ac,Braaten:1996jr,Kajantie:2002wa}.
These calculations require resummations of ordinary perturbation
theory to cope with infrared problems, which is done most elegantly
and efficiently by means of effective field theories
\cite{Braaten:1995cm}.
Up to and including order $g^6 \log(g)$ one can in fact
set up an expansion into Feynman diagrams, but this breaks
down at the order $g^6$, which receives nonperturbative contributions
from the chromomagnetostatic sector.
However, already the expansion up to that point suffers from
unusually poor convergence properties and a strong
dependence on the renormalization point even at temperatures many
orders of magnitude higher than $\Lambda_{\rm QCD}$,
which seems to reduce calculations of this kind to a purely
academic enterprise. The perturbative
expansion appears to become reliable only for coupling constants so small
that the thermodynamic potential is anyway very well approximated
by the Stefan-Boltzmann result for a noninteracting plasma.

However, this problem is not specific
to QCD at high temperature with its
nonperturbative magnetostatic sector.
Similarly poor convergence behaviour appears also in
simple scalar field theory \cite{Parwani:1995zz}, and even
in the case of large-$N$ $\phi^4$ theory \cite{Drummond:1997cw},
where all interactions can be resummed in a local thermal mass term.
As soon as one starts
expanding out in a series of powers and logarithms of the coupling,
the result for the thermodynamic potential goes wild, whereas
the exact result that is available for the large-$N$ $\phi^4$ theory
is completely unspectacular and smooth.

\begin{figure}
\centerline{\includegraphics[viewport=50 200 540 555,width=.6\textwidth]{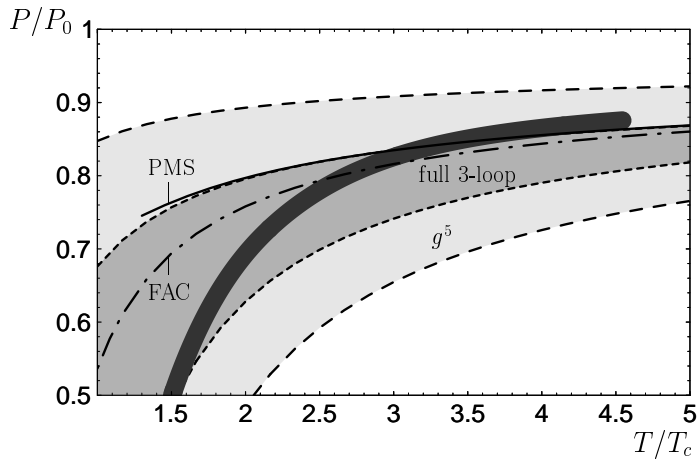}}
\caption{Three-loop pressure in pure-glue QCD 
normalized to the ideal-gas value
with 
unexpanded effective-field-theory parameters 
when $\bar\mu$ is varied between $\pi T$ and $4\pi T$ (medium-gray band).
The broad light-gray band underneath is the strictly perturbative result to
order $g^5$ with the same scale variations. The full line gives the
result upon extremalization (PMS) with respect to $\bar\mu$ (which does
not have solutions below $\sim 1.3T_c$); the dash-dotted line corresponds
to fastest apparent convergence (FAC) in $m_E^2$, which sets
$\bar\mu\approx 1.79\pi T$. (From \cite{Blaizot:2003iq})
\label{fig:3loop}}
\end{figure}

Various techniques have been developed to restore the convergence
of the perturbative series. The more physically (rather
than mathematically) motivated ones
include ``screened perturbation theory'' 
\cite{Karsch:1997gj,Andersen:2000yj},
its generalization to gauge theories (``HTL perturbation theory'')
\cite{Andersen:1999fw,Andersen:2002ey,Andersen:2003zk,Andersen:2004fp},
and the use of 2PI techniques which put the emphasis on
a quasiparticle description
\cite{Blaizot:1999ip,Blaizot:1999ap,Blaizot:2000fc,Blaizot:2003tw}.
In particular the latter results agree very well with available
lattice data down to temperatures about three times the
phase transition temperature so that the range of applicability
of weak-coupling results does not appear to be restricted to
uninterestingly high temperatures.
Moreover, these results also agree well in this regime with the perturbative
results of Ref.~\cite{Braaten:1996jr,Kajantie:2002wa} to order $g^5$ or
even $g^6 \log(g)$ provided the effective-field-theory parameters
used in this approach are not treated strictly perturbatively, but
are kept in the form in which they appear naturally
\cite{Blaizot:2003iq}.
The difference this makes is shown in Fig.~\ref{fig:3loop}, where
the light-gray band shows the renormalization-scale dependence
(a measure of the theoretical uncertainty) to strictly order $g^5$,
and the medium-gray band shows that of the unexpanded three-loop
result, both compared to the lattice result of Ref.~\cite{Boyd:1996bx},
represented by the thick dark-gray curve. In the unexpanded three-loop
result, the renormalization scale dependence is highly nonlinear
such that an extremum is attained at the upper boundary marked
PMS for principal of minimal sensitivity\footnote{A slightly different
implementation of the PMS to the thermodynamical potential
with roughly comparable results has recently been presented
in Ref. \cite{Inui:2005gu}.}; a different optimization of the perturbative
result is fastest apparent convergence (FAC), which turns out to
be close to the former.

In the following, we shall first consider the extension
to nonzero quark chemical potential
of the perturbative results obtained by Vuorinen
by means of the effective field theory
provided by the technique of dimensional reduction.
After a comparison with lattice results on the one hand, and
analytical results from the large-$N_f$ limit of QCD on the other hand,
we shall turn to the case of chemical potentials much larger than
the temperature. This regime turns out to be much richer
than the one at high temperature. For sufficiently low
temperature, new phases with colour superconductivity occur
which can in fact be analysed by weak-coupling techniques
applied to full QCD \cite{Son:1998uk,Schafer:1999jg,Pisarski:1999bf,Pisarski:1999tv,Brown:1999aq,Brown:2000eh,Wang:2001aq}. In this regime,
and also in the normal phase at temperatures above the
critical one, non-Fermi-liquid effects play a crucial role.
As we shall describe, the relevant effective field theory
in this situation is those of hard dense loops (HDL)
\cite{Altherr:1992mf,Manuel:1996td}, and
we have used it to calculate systematically non-Fermi-liquid
effects in entropy and specific heat as well as in
the dispersion laws of fermionic quasiparticles at low
temperature and high quark chemical potential.

\section{Dimensional reduction at high temperature}

When the coupling constant is small, $g\ll1$, and temperature
$T$ is the largest scale, there is a natural
hierarchy of scales:
the ``hard'' scale $T$, the ``soft'' scale $gT$, ``ultrasoft'' $g^2T$, \ldots.
In the imaginary-time formalism,
the hard scale $T$ is carried by nonzero Matsubara frequencies
$\omega_n=\pi i n T$ with $n$ even/odd for bosons/fermions,
and all the softer scales involve only the zero modes of the bosons.
The ``soft'' scale $gT$ is
where collective phenomena such as Debye screening and Landau
damping occur%
\footnote{In plasmas with a momentum-space
anisotropy, i.e.\ temperatures that depend on the direction,
the scale $g\,{\rm max}(T)$ is also the scale of
magnetic instabilities 
\cite{Mrowczynski:2000fp,Romatschke:2003ms,Arnold:2003rq,Mrowczynski:2004kv}.} 
and its effective field theory is Yang-Mills theory reduced to
three spatial dimensions\footnote{The relevant effective theory
for time-dependent problems is still 3+1-dimensional and is
given to leading order by the so-called hard-thermal-loop (HTL)
effective action \cite{Frenkel:1990br,Braaten:1990mz,Taylor:1990ia,Braaten:1992gm}.}
\be
{\cal L}_E=\2 \tr F_{ij}^2 + \tr [D_i,A_0]^2
+m_E^2 \tr A_0^2 + \2 \lambda_E (\tr A_0^2)^2 
+\ldots
\ee
In lowest order
one has a dimensionful
coupling $g_E^2 = g^2T+O(g^4)$ and \cite{Nadkarni:1988fh}
\be\label{LQCDparam}
m_E^2=(1+N_f/6)g^2 T^2+g^2\sum_q {\mu_q^2\02\pi^2}+O(g^4),\qquad
\lambda_E={9-N_f\012\pi^2}g^4T+\ldots.
\ee

The dominant contributions to the thermodynamic pressure
comes from the hard modes, and these contributions are
completely perturbative,
\be
P^{\rm hard}=T^4(c_1+c_2
g^2+c_3 g^4+c_4 g^6+\ldots),
\ee
though the coefficients $c_{\ge3}$ depend on the cutoff ($\Lambda_E$)
required to separate hard from soft scales.
The soft contributions, on the other hand, are determined by
the effective three-dimensional theory, and to three-loop order
the result only involves the parameters $g_E$ and $m_E$
\cite{Braaten:1996jr}
\be\label{P3s}
P_{\rm soft}/T = {2\03\pi}m_E^3-{3\08\pi^2}\left(
4\ln{\Lambda_E\02m_E}+3\right)g_E^2 m_E^2
-{9\08\pi^3}\left({89\024}-{11\06}\ln2+{1\06}\pi^2\right)
g_E^4\,m_E^{\phantom4}+\ldots
\ee
The complete three-loop pressure of QCD is obtained by adding
$P={P^{\rm hard}}+{P^{\rm soft}}$, and to achieve the
maximal perturbative accuracy $P^{\rm hard}$
as well as the effective field theory
parameters $g_E$ and $m_E$ are required to order $g^4$.

As long as $T\gg m_E$, this program can be extended to finite
quark chemical potential. Besides modifying the parameters
of the effective theory, there are also new, $C$-odd
terms in the effective Lagrangian. The one with smallest
dimension in nonabelian theories reads 
\cite{KorthalsAltes:1999cp,Hart:2000ha,Bodeker:2001fs}
\be\label{muA03}
\mathcal L_E^{(\mu)}=i{g^3\03\pi^2}\sum_q \mu_q \tr A_0^3.
\ee
In general the effects of these additional $C$-odd terms are
small compared to the $C$-even operators.

One quantity which is determined to leading order by
the operator (\ref{muA03}) is the flavour off-diagonal
quark number susceptibility 
at zero chemical potential \cite{Blaizot:2001vr}
\be
\chi_{ij}\equiv{\6^2 {P} \0 \6 \mu_i\6 \mu_j}.
\ee
When quark masses are negligible, all off-diagonal components
are equal at $\mu_i=0$. Denoting them by $\tilde\chi$, the
leading-order term involves a logarithmic term coming from
the exchange of three electrostatic gluons and is given
by \cite{Blaizot:2001vr}
\be\label{tcc0}
{\tilde \chi}\simeq-
{(N^2-1)(N^2-4)\0 384 N}\left({g\0\pi}\right)^6 T^2\ln{1\0g}.
\ee
where $N$ is the number of colours. This vanishes in SU(2)
gauge theory, but not in QCD, and also not in QED,
where (in the ultrarelativistic limit) \cite{Blaizot:2001vr}
\be
\tilde \chi\Big|_{\mathrm{QED}} \simeq- {e^6\024 \pi^6} T^2\ln{1\0e} . 
\ee

There has been for some time a discrepancy of the
result (\ref{tcc0}) with lattice results
on off-diagonal quark-number susceptibilities.
The authors of
Refs.~\cite{Gavai:2001ie,Gavai:2002kq} have obtained
results in the deconfined phase
that were far below those predicted by perturbation theory,
and have interpreted this as new evidence for nonperturbative
physics and the failure of weak-coupling methods. 
Most recent lattice results \cite{Bernard:2004je,Gavai:2004sd,Allton:2005gk}
have disproved the previous ones, and there is now
agreement with the perturbative estimate at $T\ge 2 T_c$.

\begin{figure}
\centerline{\includegraphics[viewport=50 240 530 540,width=.6\textwidth]{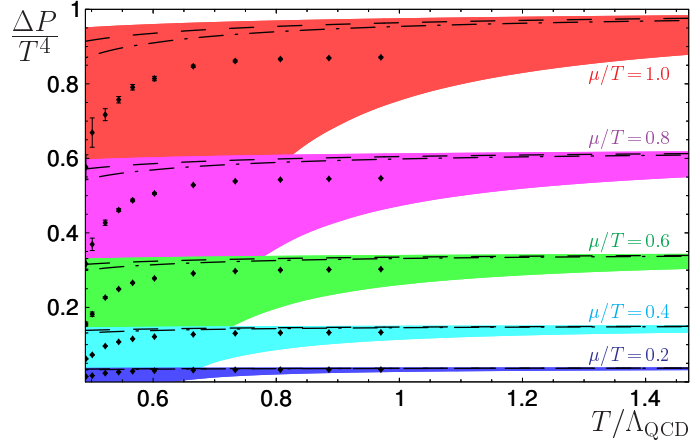}}
\caption{The difference $\Delta P=P(T,\mu)-P(T,0)$ divided by $T^4$ using
the unexpanded three-loop result from dimensional reduction
of Ref.~\cite{Vuorinen:2003fs} for $\mu/T=0.2,\ldots,1.0$
(bottom to top).
Shaded areas correspond to a variation
of $\muMS$ around the FAC-m choice by a factor of 2; dashed and
dash-dotted lines
correspond to the FAC-g and FAC-m results, respectively. 
Also included are the recent lattice data of Ref.~\cite{Allton:2003vx}
(not yet continuum-extrapolated!) assuming $T_c^{\mu=0}=
0.49\, \Lambda_{\rm QCD}$. (From \cite{Ipp:2003yz})}
\label{fig:IRV}
\end{figure}

The complete three-loop pressure of QCD at finite quark
chemical potential was recently calculated by Vuorinen \cite{Vuorinen:2003fs}.
Like the result at zero quark chemical potential,
there is poor convergence and large renormalization scale dependence
at realistic couplings, but again the apparent convergence can be
improved importantly by applying the prescription of Ref.~\cite{Blaizot:2003iq}
which keeps effective-field-theory parameters unexpanded.
Figure~\ref{fig:IRV} shows the result for
$\Delta P=P(T,\mu)-P(T,0)$ for $N_f=2$ at several values of $\mu/T$ for
which there are recent lattice data \cite{Allton:2003vx}.
The shaded regions correspond to variations of
$\muMS$ by a factor of 2 around a FAC value, and the
dashed lines correspond to (two variants of) FAC $\muMS$.
At $T/T_0=2$, the highest value considered in \cite{Allton:2003vx},
the FAC results exceed the not-yet-continuum-extrapolated 
lattice data consistently by $\approx 10\%$, which is
roughly the expected discretization error
\cite{Karsch:2000ps}.
When normalized to the free value $\Delta P_0$ instead of $T^4$, the results
would be essentially $\mu$-independent and thus also very similar to
the 
$N_f=2+1$ lattice
results of Ref.~\cite{Fodor:2002km} 
as well as the quasiparticle model
results of Refs.~\cite{Szabo:2003kg,Rebhan:2003wn}.

\section{Large-$N_f$ limit of QCD and QED}

\begin{figure}
\centerline{\includegraphics[width=.5\textwidth]{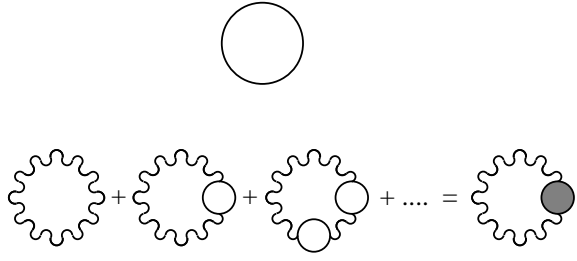}}
\caption{Diagrams contribution to the thermodynamic potential
of large-$N_f$ QCD}
\label{fig:lnf}
\end{figure}

While the comparisons of weak-coupling results
with lattice data are in remarkably good shape at temperatures
a few times the deconfinement temperature, it seems desirable
to have a cleaner testing ground for the resummation procedures
required in applying weak-coupling techniques at realistic
temperatures. Guy Moore \cite{Moore:2002md} has proposed to use the
large-$N_f$ limit of QCD and QED for this purpose.
Large-$N_f$ QCD is much simpler than large-$N_c$ QCD.
In the limit $N_f\to\infty$, $N_c\sim 1$,
$g^2 N_f \sim 1$, the relevant diagrams are those displayed
in Fig.~\ref{fig:lnf}. They involve a dressed gluon propagator
which contains typical gauge-theory phenomena such as 
Debye screening for electrostatic modes, 
unscreened magnetostatic modes,
complicated dispersion laws, Landau damping, and also plasmon damping.
This is therefore a much richer theory than the large-$N$ scalar
field theories that are frequently
used for testing resummations of thermal perturbation theory.

In terms of the polarization tensor $\Pi^{\mu\nu}=\Pi^{\mu\nu}_{\rm vac}+
{\Pi^{\mu\nu}_{\rm mat}}$, where the matter part $\Pi^{\mu\nu}_{\rm mat}$
can be decomposed in a transverse and a spatially longitudinal
piece, $\Pi_T$ and $\Pi_L$, respectively, the thermal pressure
reads
\bea\label{lnfP}
P & = & NN_f\left( {7\pi^2 T^4\0180}+{\mu^2 T^2\06}+{\mu^4\012\pi^2} \right)
\nonumber\\&+&N_g
\int \frac{d^{3}q}{(2\pi )^{3}}\int _{0}^{\infty }\frac{d\omega }{\pi }
\biggr[ \,2\,\Bigl\{ \left[n_{b}+{\textstyle\frac{1}{2}}\right]\,
\rm {Im}\ln \left(q^{2}-\omega ^{2}+{\Pi _{T}}+\Pi _{\rm {vac}}\right)
\nonumber\\
 &  & \qquad \qquad \qquad \qquad \quad 
-{\textstyle\frac{1}{2}}\,\rm {Im}
\ln \left(q^{2}-\omega ^{2}+\Pi _{\rm {vac}}\right)\Bigr\} \nonumber\\
 &  &  +\,\Bigl\{ \left[n_{b}+{\textstyle\frac{1}{2}}\right]\,
\rm {Im}\ln \frac{q^{2}-\omega ^{2}+{\Pi _{L}}+
\Pi _{\rm {vac}}}{q^{2}-\omega ^{2}}-{\textstyle\frac{1}{2}}\,
\rm {Im}\ln \frac{q^{2}-\omega ^{2}
+\Pi _{\rm {vac}}}{q^{2}-\omega ^{2}}\Bigr\} \biggr] +O(N_f^{-1}).
\eea

The effective coupling is hidden in $\Pi_{L,T}$ and given by
\be
\g^2 = \left\{
        \begin{array}{cc} 
        \displaystyle \frac{g^2 N_f}{2} \, , & {\rm QCD} \, , \\ & \\
        e^2 N_f \, , & {\rm QED} \, . \\ \end{array} \right.
\ee
with exact one-loop beta function:
\be
\frac{1}{g_{\rm {eff}}^{2}(\mu )}=\frac{1}{g_{\rm {eff}}^{2}(\mu ')}
+\frac{\ln (\mu '/\mu )}{6\pi ^{2}}\,.
\ee
There is no asymptotic freedom. In fact, there is a Landau
singularity at exponentially large scales $\Lambda _{\rm L}=\bar\mu_{\rm MS}
 e^{5/6}e^{6\pi ^{2}/\g^{2}(\bar\mu_{\rm MS})},$
implying this theory exists only as a cutoff theory with
$\Lambda_{\rm Cutoff}<\Lambda_L$. However, for the purpose
of testing thermodynamic results this is no problem
as long as the effective cutoff provided by temperature and/or
chemical potential $T,\mu \ll \Lambda _{\rm L}$.
It does however lead to certain technical intricacies, since
$\Lambda_{\rm Cutoff}$ needs to be implemented such that
Euclidean invariances are respected in order not to produce
spurious singularities \cite{Moore:2002md,Ipp:2003zr}.

\begin{figure}
\begin{center}\includegraphics[width=.7\textwidth]{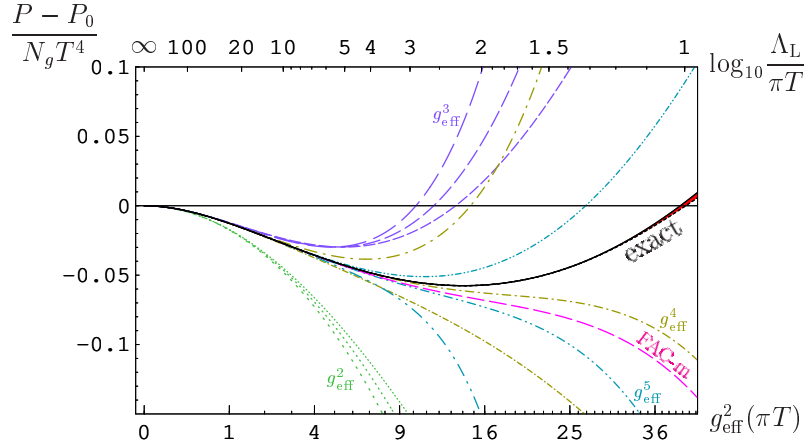}\end{center}

\caption{Exact result for the interaction pressure at finite temperature and
zero chemical potential at large $N_{f}$ as a function of $\geff^{2}(\muMS=\pi T)$,
compared to strict perturbation theory \cite{Moore:2002md,Ipp:2003zr,Ipp:2003jy}.
The tiny red band appearing for large values of the coupling for the exact
result shows the cutoff dependence from varying the upper numerical
integration cutoff between 50\% and 70\%
of the Landau pole $\Lambda_{L}$. The results of strict perturbation
theory are given through order $\geff^{2}$ (dotted line), $\geff^{3}$
(dashed), $\geff^{4}$ (dash-dotted), and $\geff^{5}$ (dash-dot-dotted)
where the renormalization scale $\muMS$ is varied between $\frac{1}{2}\pi T$
(line pattern slightly compressed), $\pi T$, and $2\pi T$ (line
pattern slightly stretched). The line labelled {}``FAC-m'' indicates
the scale chosen by the prescription of fastest apparent convergence
for which the curves of $\geff^{4}$ and
$\geff^{5}$ coincide. (From \cite{Blaizot:2005fd}) \label{fig:lnfpvspt}}
\end{figure}

Comparing the available perturbative results with the ``exact''
large-$N_f$ result obtained numerically \cite{Moore:2002md,Ipp:2003zr}
shows that strict perturbation theory has the usual
problems of poor convergence and large $\bar\mu$-dependences
(Fig.~\ref{fig:lnfpvspt}). The exact result has a curious nonmonotonic
behaviour as a function of $\g^2$, and only the small deviations
from the free pressure at small coupling seem to be unambiguously
predicted by perturbation theory.

\begin{figure}
\centerline{\includegraphics[width=.7\textwidth]{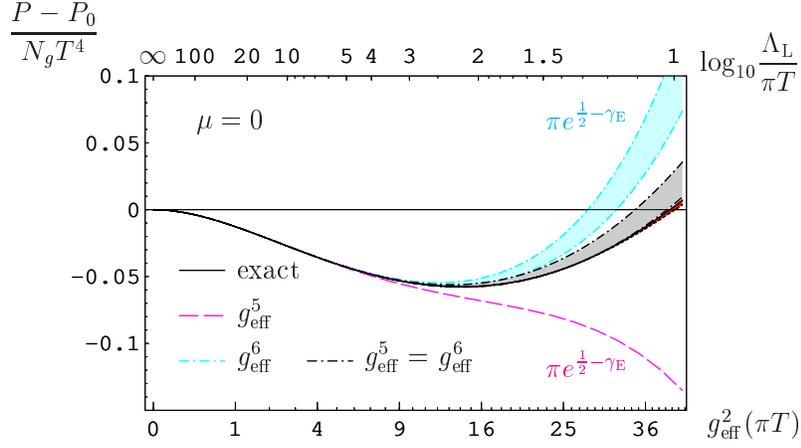}}
\caption{\label{fig:zeromu}
Exact result for the interaction pressure of large-$N_f$ QCD at zero
chemical potential \cite{Moore:2002md,Ipp:2003zr}
as a function of $\g^2(\bar\mu_{\rm MS}=\pi T)$ or, alternatively,
$\log_{10}(\Lambda_{\rm L}/\pi T)$. The purple dashed line is
the perturbative result when the latter is evaluated
with renormalization scale
$\bar\mu_{\rm MS}=\bar\mu_{\rm FAC}\equiv\pi e^{1/2-\gamma}T$;
the blue dash-dotted lines include the numerically determined
coefficient to order $\g^6$ (with its estimated error)
at the same renormalization scale.
The result marked ``$\g^5=\g^6$'' 
corresponds to choosing $\bar\mu_{\rm MS}$ such
that the order-$\g^6$ coefficient vanishes and retaining
all higher-order terms contained in the plasmon term $\propto m_E^3$.
}
\end{figure}

Figure \ref{fig:zeromu} shows the result of an optimized $g^6$
result with the prescription of keeping the effective-field-theory
parameter $m_E$ unexpanded. (The coefficient of the $g^6$ term
has been extracted numerically with a 10\% error shown by shaded
bands.) The result agrees exceedingly well with the exact result
even for the largest coupling where the ambiguity introduced
by the Landau singularity is still under control.

Perhaps even more remarkably, the nontrivial behaviour of the
thermodynamic potential has recently been reproduced with
comparable accuracy by using a 2-loop $\Phi$-derivable
approximation to the entropy (where the perturbative accuracy
is actually only order $g^3$) \cite{Blaizot:2005wr}.

\begin{figure}
\centerline{\includegraphics[width=.7\textwidth]{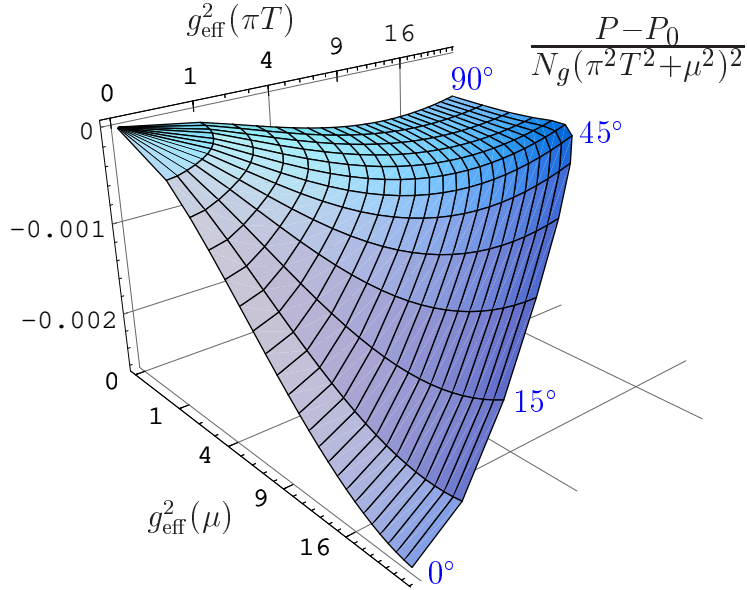}}
\caption{\label{fig:3d}
Exact result for the large-$N_f$ interaction pressure $P-P_0$ normalized
to $N_g(\pi^2T^2+\mu^2)^2$ as a function of $\g^2(\bar\mu_{\rm MS})$
with $\bar\mu_{\rm MS}^2=\pi^2T^2+\mu^2$, which is the radial
coordinate, and $\varphi=\arctan(\pi T/\mu)$.}
\end{figure}

In Ref.~\cite{Ipp:2003jy} two of us have also evaluated the
large-$N_f$ pressure at finite chemical potential, which is displayed
in the 3-d plot of Fig.~\ref{fig:3d}.

\begin{figure}
\centerline{\includegraphics[width=.6\textwidth]{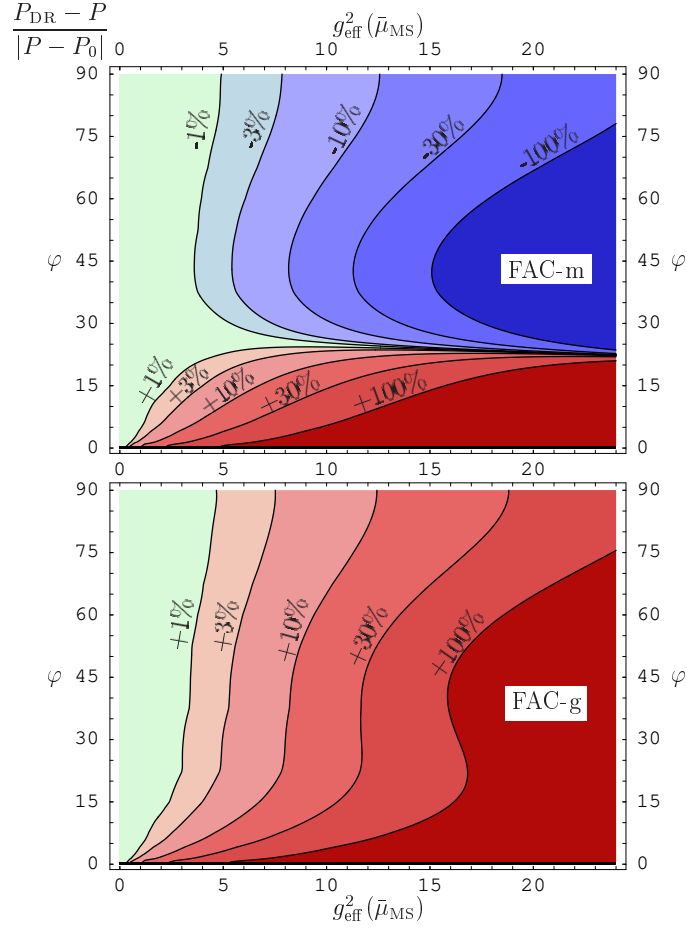}}
\caption{Percentage errors of the perturbative result for the
interaction part of the pressure to
order $\geff^5$ in the large-$N_f$ limit
for two choices of $\muMS$: Fastest apparent convergence
of $P$ as well as
$m_E^2$ (FAC-m), and of $g_E^2$ (FAC-g).
The brightest area corresponds to an error of less than 1\%,
the darkest ones to an error of over 100\%.
The ratio of chemical potential to temperature increases from
top to bottom according to $\varphi=\arctan(\pi T/\mu)$ so that
$90^\circ$ corresponds to high temperature and zero chemical
potential, and $0^\circ$ to zero temperature and high chemical potential.
The coupling is given in terms of
$\geff^2(\muMS)$ at $\muMS=\sqrt{\pi^2 T^2+\mu^2}$. (From \cite{Ipp:2003yz})
\label{fig:contpl}}
\end{figure}

In Fig.~\ref{fig:contpl} this result is compared with the
perturbative result to order $g^5$ for two optimized renormalization
scales. While this shows a fairly large region where
the perturbative results obtained through dimensional reduction
is a good approximation, it also shows that there is
a breakdown of dimensional reduction at very small temperature
$T<0.1\mu$.
In this regime it becomes necessary to evaluate all
infrared sensitive diagrams such as ring diagrams
in four dimensions. Combining such procedures with
the existing analytical results to three-loop order
should be able to fill in the existing gap
of perturbative results on the pressure for very small temperatures
and high chemical potential \cite{IKRV}.

A three-loop result for the QCD pressure does exist at exactly
zero temperature. It is due to Freedman and McLerran 
\cite{Freedman:1977dm,Freedman:1977ub,Baluni:1978ms}
and it has recently been updated by Vuorinen \cite{Vuorinen:2003fs}.
In the large-$N_f$ limit it reads
\be
{P-P_0\0N_g\,\mu^4 }\Big|_{T=0}=
-{\g^2\032\pi^4}-\left[\ln{\g^2\02\pi^2}-{2\03}\ln{\bar\mu_{\rm MS}\0\mu}
-0.53583\ldots\right]
{\g^4\0128\pi^6}+O(\g^6\ln\g^{\phantom{6}}).
\ee

The exact (but numerical) large-$N_f$ result of Ref.~\cite{Ipp:2003jy}
in fact provides a check of these results as well as a numerical
determination of the next terms to order $\g^6$ and $\g^6 \ln(\g)$:
\be
{P-P_0\0N_g\,\mu^4 }\Big|_{\g^6, T=0, \bar\mu_{\rm MS}=\mu}
=[3.18(5) \ln{2\pi^2\0\g^2}+3.4(3)]{\g^6\02048\pi^8}.
\ee

\begin{figure}
\centerline{\includegraphics[width=.7\textwidth]{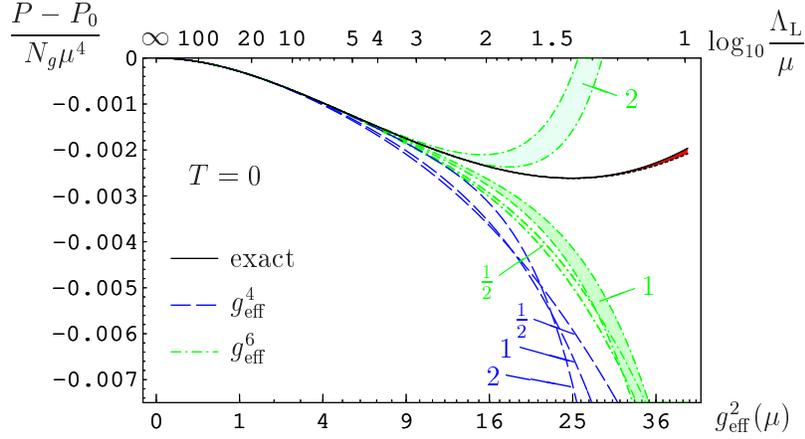}}
\caption{\label{fig:zerot}
The interaction part of the pressure at zero temperature and
finite chemical potential as a function of $\g^2(\bar\mu_{\rm MS}=\mu)$
or, alternatively, $\log_{10}(\Lambda_{\rm L}/\mu)$,
compared with the perturbative result of Freedman and McLerran
\cite{Freedman:1977dm,
Baluni:1978ms}
to order $\g^4$, and our numerically extracted order-$\g^6$
result, both with renormalization scale in the perturbative
results varied around $\bar\mu_{\rm MS}=\mu$ by a factor of 2.
The coloured bands of the $\g^6$-results
cover the error of the numerically extracted perturbative coefficients.
(From \cite{Ipp:2003jy})}
\end{figure}

Figure \ref{fig:zerot} compares the exact and the perturbative
results for the large-$N_f$ pressure at zero temperature and
high chemical potential, with the perturbative results
evaluated for three values of $\muMS/\mu$.

\section{Non-Fermi-liquid behavior}

From the leading-order interaction term in the thermal pressure
\cite{Kap:FTFT}
\be\label{P2pt}
P-P_0=- N_g\left[ {5\09} T^4 + {2\0\pi^2} \mu^2 T^2 +{1\0\pi^4}\mu^4 
\right]{\g^2\032} + \ldots
\ee
one would expect that the entropy density $\mathcal S = \left(
{\6 P\0\6 T} \right)_\mu$ at small $T\ll\mu$ should start as
\be
\mathcal S -\mathcal S_0={-}N_g {\g^2\08\pi^2} \mu^2 T + \ldots
\ee

\begin{figure}
\centerline{\includegraphics[width=.7\textwidth]{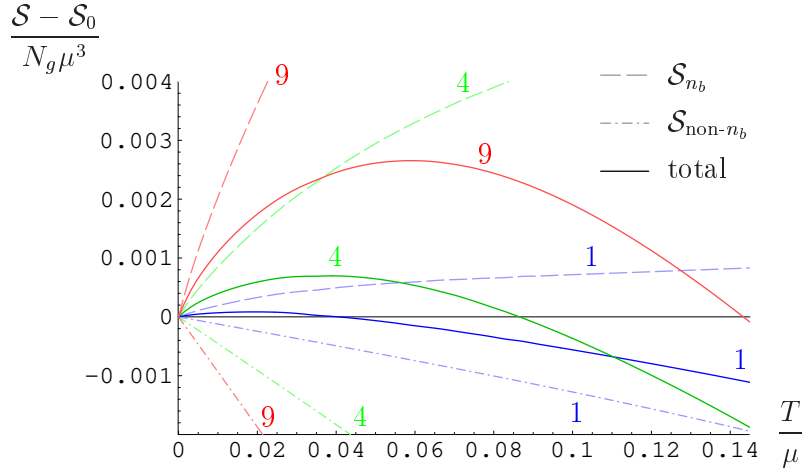}}
\caption{\label{fig:sst}
The interaction part of the
entropy at small $T/\mu$ for $\g^2(\bar\mu_{\rm MS}\!=\!\mu)=1$, $4$,
and $9$.
The ``non-$n_b$'' contributions (dash-dotted lines)
are negative and approximately linear
in $T$ with a coefficient agreeing with the exchange term $\propto \g^2$
in the pressure at small coupling; the ``$n_b$'' contributions 
(dashed lines), which are
dominated by transverse gauge boson modes, are
positive and nonlinear in $T$ such that the total entropy exceeds
the free-theory value at sufficiently small $T/\mu$.
}
\end{figure}

However, for sufficiently small $T$, namely when $T\ll g\mu$,
the interaction part of the entropy density 
extracted from the exact large-$N_f$ result (Fig.~\ref{fig:sst})
has even a different sign. The contributions to (\ref{lnfP})
which do not involve the Bose distribution function $n_b$
do indeed behave in accordance to the result (\ref{P2pt}),
but those with the function $n_b$ that one would expect
to be negligible at very small temperature turn out to
become the dominant ones, and they contribute terms
which are nonlinear in $T$, implying a strong deviation
from the usual behaviour of Fermi liquids \cite{LL:V}.
This non-Fermi-liquid behaviour is caused by long-range
interactions, namely weakly screened magnetic modes.
At very small frequencies, the latter have a dynamical
screening length $\kappa = [\pi m_{\rm D}^2\omega/4]^{1/3}$
and it has been found long ago by Holstein, Norton, and Pincus
\cite{Holstein:1973} that this 
manifests itself in 
the appearance of an anomalous contribution to the 
low-temperature limit of entropy and specific
heat proportional to $\alpha T\ln T^{-1}$.

While this effect is perhaps too small for experimental
detection in nonrelativistic situations, it drew renewed theoretical attention more recently
\cite{Reizer:1989,Gan:1993,Chakravarty:1995}
after the detection of non-Fermi-liquid behavior in
the normal state of high-temperature superconductors \cite{Varma:1989}
and in other systems of strongly correlated electrons,
which may be due to effective gauge field dynamics
(see also \cite{Polchinski:1992ed,Polchinski:1994ii,Nayak:1994ng}).

In deconfined degenerate quark matter, the analogous effect
can more easily be important because the larger
coupling constant $\alpha_s$ together with the relatively 
large number of gauge bosons
increases the numerical value of the effect by orders of magnitude.
In contrast to the case of a high-temperature quark-gluon plasma,
chromomagnetostatic fields are expected to remain unscreened in the
low-temperature limit \cite{Son:1998uk} and thus lead to the same singularities
in the fermion self-energy that are responsible for the
breakdown of the Fermi-liquid description in the nonrelativistic
electron gas considered in \cite{Holstein:1973}.

An important consequence of such non-Fermi-liquid behavior in
quantum chromodynamics (QCD) is a reduction of the magnitude of
the gap in color superconductors \cite{Son:1998uk,Brown:1999aq,Brown:2000eh,%
Wang:2001aq} which on the basis of weak-coupling calculations
are estimated to have a critical temperature 
in the range between 6 and 60 MeV \cite{Rischke:2003mt}.
Quark matter above this temperature
has long-range chromomagnetic interactions that
should lead to an anomalous specific heat with possible
relevance for the cooling of young neutron stars 
as pointed out by Boyanovsky and de Vega 
\cite{Boyanovsky:2000bc,Boyanovsky:2000zj}.
However, in Ref.~\cite{Boyanovsky:2000zj} these authors
claimed that the $\alpha T \ln T^{-1}$ term in the specific heat
as reported in \cite{Holstein:1973,Gan:1993,Chakravarty:1995}
would not exist, neither in QCD nor in QED.
Instead they obtained a $\alpha T^3 \ln T$ correction
to the leading ideal-gas behavior, which by renormalization-group
arguments was resummed into a $T^{3+O(\alpha)}$ correction
as the leading non-Fermi-liquid effect on the specific heat.\footnote{%
Resummation of the $\alpha T \ln T^{-1}$ term along the
lines of Ref.~\cite{Boyanovsky:2000zj} would have
led to a $T^{1+O(\alpha)}$ term instead.}
At low temperatures, such a contribution would be rather
negligible compared to standard perturbative corrections to
the ideal-gas result $\propto T$.

In Refs.~\cite{Ipp:2003cj,Gerhold:2004tb} we have recently
been able to resolve these contradictory results and shown
that the $\alpha T \ln T^{-1}$ term as obtained in
\cite{Holstein:1973,Gan:1993,Chakravarty:1995} is indeed
correct. In addition to the coefficient of this leading logarithm,
we have calculated the coefficient under the log as well as
higher terms in the low-$T$ series, which come with fractional
powers of $T$,
\bea\label{Sseries}
&&{\mathcal S-\mathcal S_0\0N_g}= \frac{\geff ^{2}\mu^2 {T}}{36\pi ^{2}}
\left( \ln{4\g\mu\0\pi^2{T}}-2+\gamma_{E}-\frac{6}{\pi ^{2}}\zeta '(2)
\right)\nonumber\\&&\qquad- 
\frac{8\;2^{2/3}\Gamma (\frac{8}{3})\zeta (\frac{8}{3})}{9\sqrt{3}\pi ^{11/3}}(\geff \mu )^{4/3} {T^{5/3}} 
+ \frac{80\;2^{1/3}\Gamma (\frac{10}{3})\zeta (\frac{10}{3})}{27\sqrt{3}\pi ^{13/3}}(\geff \mu )^{2/3} {T^{7/3}}\nonumber\\
&&\qquad+{{2048-256\pi^2-36\pi^4+3\pi^6\0540\pi^2}}{T^3\left[\ln {{ \geff\mu}\0T}{-{4.3493485\ldots}}\right]}+
{O(T^{11/3})}.
\eea

\begin{figure}
\centerline{\includegraphics[
width=.7\textwidth]{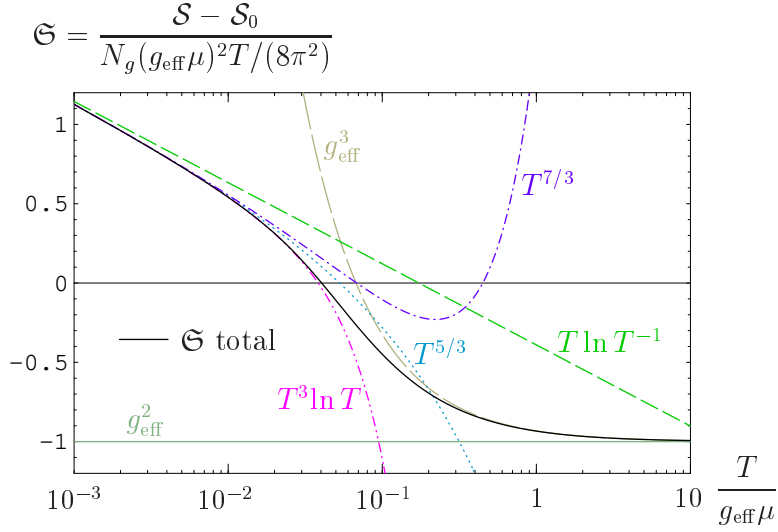}}
\caption{The function $\mathfrak S(T/(\g\mu))$
which determines the leading-order interaction contribution
to the low-temperature entropy.
The normalization is such that $\mathfrak S=-1$ corresponds
to the result of ordinary perturbation theory.
The various dash-dotted lines give the first few orders of the
low-temperature series for the entropy; the perturbative
result from dimensional reduction to order $\g^2$ and $\g^3$
is given by the full and the long-dashed curves, resp. (from \cite{Gerhold:2004tb}) \label{fig:Sfrak}}
\end{figure}

This expansion is systematic for $T\ll \g\mu$, but breaks down
at $T\sim \g\mu$; for $T\gg \g\mu$ it in fact has to switch to
the perturbative series that can be obtained from dimensional reduction.
A result which covers all $T\ll\mu$ can be obtained
by resumming the complete nonlocal HDL self energies and
is given by the compact expression \cite{Gerhold:2004tb}
\bea\label{SHDL}
{1\0N_g}(\mathcal S-\mathcal S^0)&=&-{\g^2\mu^2T\024\pi^2}
-{1\02\pi^3}\int_0^\infty dq_0\, {\6n_b(q_0)\0\6T}
\int_0^\infty dq\,q^2\,\biggl[ \nonumber\\&&\qquad
2\,\im \ln \left( q^{2}-q_{0}^{2}+\Pi^{\rm HDL} _{T} \0 q^{2}-q_{0}^{2} \right )
+\im \ln \left( \frac{q^{2}-q_{0}^{2}+\Pi^{\rm HDL} _{L}}{q^{2}-q_{0}^{2}}\right)
\biggr]\nonumber\\&& + O(\g^4\mu^2T).\;\;\quad
\eea
For $T\ll g\mu$ this resums the above low-$T$ series which
is dominated by magnetic effects and for $\g\mu \ll T \ll \mu$
connects smoothly to the dimensional reduction result
\be
{1\0N_g}(\mathcal S-\mathcal S_0)
\simeq -{\g^2\mu^2 T\08\pi^2}+{\g^3\mu^3\012\pi^4}
\ee
where the $O(g^3)$ contribution is from electrical Debye screening.
The numerical evaluation of (\ref{SHDL}) together with the
first few orders of the low-temperature series and
the perturbative result from dimensional reduction is shown
in Fig.~\ref{fig:Sfrak}.


\begin{figure}
\centerline{\includegraphics[
width=.7\textwidth]{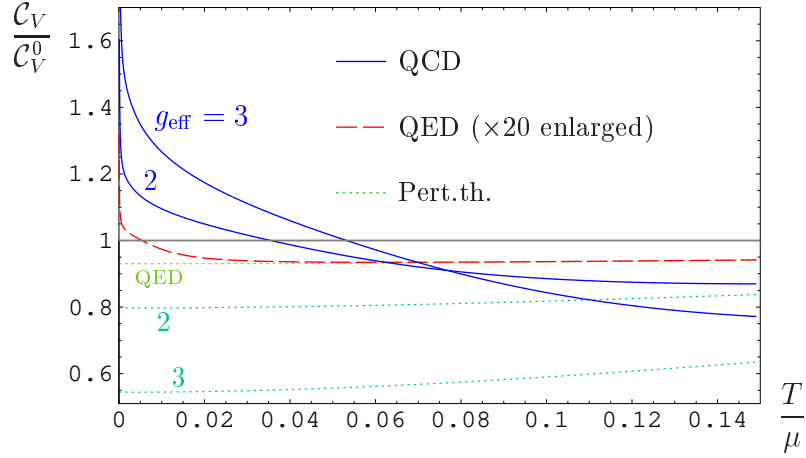}}
\caption{The HDL-resummed result for the specific heat
$\mathcal C_v$,
normalized to the ideal-gas value
for $\g=2$ and 3 corresponding
to $\alpha_s\approx 0.32$ and $0.72$ in two-flavor QCD,
and
$\g\approx 0.303$ for QED.
The deviation of the QED result from
the ideal-gas value is enlarged by a factor of 20
to make it more visible.
\label{figspecificheat}}
\end{figure}

Figure \ref{figspecificheat} displays the resulting specific
heat normalized to its ideal-gas value. 
The lines marked ``QED'' correspond to
$\g=0.303$ or $\alpha_{QED}\approx 1/137$, and the results
for $\g=2,3$ correspond to $\alpha_s\approx 0.32,0.72$
in two-flavor QCD. (Recall that $\g^2\equiv g^2N_f/2$.)
While in QED the effect is tiny (the deviations
from the ideal-gas value have been enlarged by a factor of 20
in Fig.~\ref{figspecificheat} to make them more visible), in QCD we find that
there is an interesting range of $T/\mu$ where there is
a significant excess of the specific heat over its ideal-gas value,
whereas ordinary perturbation theory \cite{Kap:FTFT}
would have resulted in 
a low-temperature limit of
$\mathcal C_v/\mathcal C_v^0=1-2\alpha_s/\pi$.
At sufficiently low temperature, $\mathcal C_v/\mathcal C_v^0$
may even become much larger than 1. As discussed in Ref.~\cite{Schafer:2004zf},
the large logarithm that appears in this limit is actually stable
against higher-order corrections. However, in QCD (though not in
QED) the growth of $\mathcal C_v/\mathcal C_v^0$ is limited
by the appearance of a superconducting phase at sufficiently small $T$.
According to Ref.~\cite{Rischke:2003mt},
the critical temperature for the color superconducting phase
transition may be anywhere between 6 and 60 MeV,
so with e.g.\ a quark chemical potential
of $\mu=500$ MeV the range $T/\mu \ge 0.012$ in Fig.~\ref{figspecificheat} 
might correspond to normal quark matter with anomalous
specific heat.
While the effect remains
small in QED, it seems conceivable that the anomalous terms in
the specific heat play a noticeable role in the thermodynamics of
proto-neutron stars,
in particular its cooling behavior in its earliest stages
before entering color superconductivity
\cite{Iwamoto:1980eb,Carter:2000xf}.

The cooling of (proto-)neutron stars with a normal quark matter
component is not only sensitive to non-Fermi-liquid effects
in the specific heat. The latter are in fact overcompensated
by two powers of $\alpha_s\ln(m/T)$ appearing
in the neutrino emissitivity \cite{Schafer:2004jp}
which come from logarithmic singularities in the
group velocities of quark quasiparticles.
While Ref.~\cite{Schafer:2004jp} assumed a single scale
for the two kinds of non-Fermi-liquid logarithms, a recent
explicit calculation \cite{Gerhold:2005uu} 
showed that these can be very different.

The scale of the logarithm in the specific heat is approximately
given by $\log(0.282m/T)$
with $m^2\equiv N_f(g\mu/2\pi)^2$. At zero temperature,
the group velocity of quark
quasiparticles near the (would-be) Fermi surface ($\varepsilon=E-\mu$)
reads \cite{Gerhold:2005uu} 
\be
v_g^{-1}(\varepsilon)=1+{g^2 C_f\012\pi^2}\ln{{8.07} m\0|\varepsilon|}
+O((\varepsilon/m)^{2/3})
\ee
which at small nonzero temperature $T\ll g\mu$ is bounded by
\be
v_g^{-1}(0)=1+{g^2 C_f\012\pi^2}\ln{{9.15}m\0T}+O((T/m)^3).
\ee
Evidently the scale of the latter logarithm differs from that
in the anomalous specific heat by more than a factor of 30.

\begin{figure}
\centerline{\includegraphics[
width=.6\textwidth]{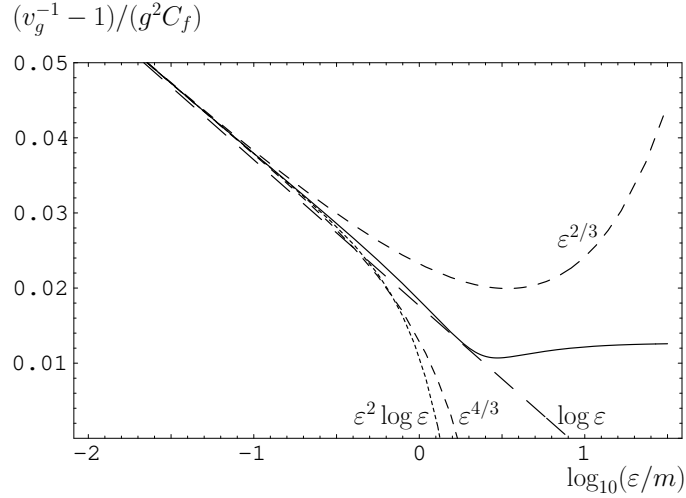}}
\caption{The group velocity of quark quasiparticles plotted as
$(v_g^{-1}-1)/(g^2C_f)$ over $\log_{10}(E-\mu/m)$
at zero temperature. 
\label{fig:invv}}
\end{figure}

Higher terms in the small $\varepsilon$ expansion of the
group velocity $v_g^{-1}(\varepsilon)$ also involve fractional
powers, and a complete result for general $\varepsilon$ can
again be obtained by HTL resummation. The latter is shown
in Fig.~\ref{fig:invv} together with the first few terms of
the small-$\varepsilon$ expansion.

\section{Conclusions}

There has recently been substantial progress in applying
weak-coupling methods to the thermodynamics of deconfined QCD,
and we have made a case that their generally poor apparent
convergence can be greatly improved by suitable resummations.
Results obtained from the effective field theory of dimensional
reduction remain predictive for temperatures down to
a few times the deconfinement temperature (rather than
many orders of magnitude higher as previously concluded).
This also holds true for nonzero chemical potential $\mu\sim T$,
and in fact comparisons with the available lattice data
show reasonable agreement already at $T\sim 2 T_c$.

For temperatures
$T\sim g\mu$ or smaller, the perturbative expansion obtained in the
dimensional reduction framework eventually breaks
down because of non-Fermi-liquid effects. Those can be calculated
analytically by HDL resummation and give rise to anomalous
terms in the specific heat and related quantities.
While tiny in QED, such effects may become important
in the thermodynamics of neutron stars with a normal
quark matter component.

\acknowledgments

This work was supported by the Austrian Science Foundation FWF,
project no.\ P16387-N08.


\begin{thebibliography}{10}

\bibitem{Arnold:1995eb}
P.~Arnold and C.-X. Zhai, {\it The three-loop free energy for high temperature
  {QED} and {QCD} with fermions},  {\em Phys. Rev.} {\bf D51} (1995)
  1906--1918, [\href{http://xxx.lanl.gov/abs/hep-ph/9410360}{{\tt
  hep-ph/9410360}}].

\bibitem{Zhai:1995ac}
C.-X. Zhai and B.~Kastening, {\it The free energy of hot gauge theories with
  fermions through {$g^5$}},  {\em Phys. Rev.} {\bf D52} (1995) 7232,
  [\href{http://xxx.lanl.gov/abs/hep-ph/9507380}{{\tt hep-ph/9507380}}].

\bibitem{Braaten:1996jr}
E.~Braaten and A.~Nieto, {\it Free energy of {QCD} at high temperature},  {\em
  Phys. Rev.} {\bf D53} (1996) 3421--3437,
  [\href{http://xxx.lanl.gov/abs/hep-ph/9510408}{{\tt hep-ph/9510408}}].

\bibitem{Kajantie:2002wa}
K.~Kajantie, M.~Laine, K.~Rummukainen, and Y.~Schr{\"o}der, {\it The pressure
  of hot {QCD} up to {$g^6\ln(1/g)$}},  {\em Phys. Rev.} {\bf D67} (2003)
  105008, [\href{http://xxx.lanl.gov/abs/hep-ph/0211321}{{\tt
  hep-ph/0211321}}].

\bibitem{Braaten:1995cm}
E.~Braaten and A.~Nieto, {\it Effective field theory approach to high
  temperature thermodynamics},  {\em Phys. Rev.} {\bf D51} (1995) 6990--7006,
  [\href{http://xxx.lanl.gov/abs/hep-ph/9501375}{{\tt hep-ph/9501375}}].

\bibitem{Parwani:1995zz}
R.~Parwani and H.~Singh, {\it The pressure of hot {$g^2\phi^4$} theory at order
  {$g^5$}},  {\em Phys. Rev.} {\bf D51} (1995) 4518--4524,
  [\href{http://xxx.lanl.gov/abs/hep-th/9411065}{{\tt hep-th/9411065}}].

\bibitem{Drummond:1997cw}
I.~T. Drummond, R.~R. Horgan, P.~V. Landshoff, and A.~Rebhan, {\it Foam diagram
  summation at finite temperature},  {\em Nucl. Phys.} {\bf B524} (1998)
  579--600, [\href{http://xxx.lanl.gov/abs/hep-ph/9708426}{{\tt
  hep-ph/9708426}}].

\bibitem{Karsch:1997gj}
F.~Karsch, A.~Patk{\'o}s, and P.~Petreczky, {\it Screened perturbation theory},
   {\em Phys. Lett.} {\bf B401} (1997) 69--73,
  [\href{http://xxx.lanl.gov/abs/hep-ph/9702376}{{\tt hep-ph/9702376}}].

\bibitem{Andersen:2000yj}
J.~O. Andersen, E.~Braaten, and M.~Strickland, {\it Screened perturbation
  theory to three loops},  {\em Phys. Rev.} {\bf D63} (2001) 105008,
  [\href{http://xxx.lanl.gov/abs/hep-ph/0007159}{{\tt hep-ph/0007159}}].

\bibitem{Andersen:1999fw}
J.~O. Andersen, E.~Braaten, and M.~Strickland, {\it Hard-thermal-loop
  resummation of the free energy of a hot gluon plasma},  {\em Phys. Rev.
  Lett.} {\bf 83} (1999) 2139--2142,
  [\href{http://xxx.lanl.gov/abs/hep-ph/9902327}{{\tt hep-ph/9902327}}].

\bibitem{Andersen:2002ey}
J.~O. Andersen, E.~Braaten, E.~Petitgirard, and M.~Strickland, {\it {HTL}
  perturbation theory to two loops},  {\em Phys. Rev.} {\bf D66} (2002) 085016,
  [\href{http://xxx.lanl.gov/abs/http://arXiv.org/abs/hep-ph/0205085}{{\tt
  http://arXiv.org/abs/hep-ph/0205085}}].

\bibitem{Andersen:2003zk}
J.~O. Andersen, E.~Petitgirard, and M.~Strickland, {\it Two-loop {HTL}
  thermodynamics with quarks},  {\em Phys. Rev.} {\bf D70} (2004) 045001,
  [\href{http://xxx.lanl.gov/abs/hep-ph/0302069}{{\tt hep-ph/0302069}}].

\bibitem{Andersen:2004fp}
J.~O. Andersen and M.~Strickland, {\it Resummation in hot field theories},
  {\em Ann. Phys.} {\bf 317} (2005) 281--353,
  [\href{http://xxx.lanl.gov/abs/hep-ph/0404164}{{\tt hep-ph/0404164}}].

\bibitem{Blaizot:1999ip}
J.~P. Blaizot, E.~Iancu, and A.~Rebhan, {\it The entropy of the {QCD} plasma},
  {\em Phys. Rev. Lett.} {\bf 83} (1999) 2906--2909,
  [\href{http://xxx.lanl.gov/abs/hep-ph/9906340}{{\tt hep-ph/9906340}}].

\bibitem{Blaizot:1999ap}
J.~P. Blaizot, E.~Iancu, and A.~Rebhan, {\it Self-consistent hard-thermal-loop
  thermodynamics for the quark-gluon plasma},  {\em Phys. Lett.} {\bf B470}
  (1999) 181--188, [\href{http://xxx.lanl.gov/abs/hep-ph/9910309}{{\tt
  hep-ph/9910309}}].

\bibitem{Blaizot:2000fc}
J.~P. Blaizot, E.~Iancu, and A.~Rebhan, {\it Approximately self-consistent
  resummations for the thermodynamics of the quark-gluon plasma: Entropy and
  density},  {\em Phys. Rev.} {\bf D63} (2001) 065003,
  [\href{http://xxx.lanl.gov/abs/hep-ph/0005003}{{\tt hep-ph/0005003}}].

\bibitem{Blaizot:2003tw}
J.~P. Blaizot, E.~Iancu, and A.~Rebhan, {\it Thermodynamics of the
  high-temperature quark-gluon plasma},  in {\em Quark-Gluon Plasma 3} (R.~C.
  Hwa and X.-N. Wang, eds.).
\newblock World Scientific, Singapore, 2003.

\bibitem{Blaizot:2003iq}
J.~P. Blaizot, E.~Iancu, and A.~Rebhan, {\it On the apparent convergence of
  perturbative {QCD} at high temperature},  {\em Phys. Rev.} {\bf D68} (2003)
  025011, [\href{http://xxx.lanl.gov/abs/hep-ph/0303045}{{\tt
  hep-ph/0303045}}].

\bibitem{Boyd:1996bx}
G.~Boyd, J.~Engels, F.~Karsch, E.~Laermann, C.~Legeland, M.~L{\"u}tgemeier, and
  B.~Petersson, {\it Thermodynamics of {SU(3)} lattice gauge theory},  {\em
  Nucl. Phys.} {\bf B469} (1996) 419--444,
  [\href{http://xxx.lanl.gov/abs/hep-lat/9602007}{{\tt hep-lat/9602007}}].

\bibitem{Inui:2005gu}
M.~Inui, A.~Niegawa, and H.~Ozaki, {\it Improvement of the hot {QCD} pressure
  by the minimal sensitivity criterion},
  \href{http://xxx.lanl.gov/abs/hep-ph/0501277}{{\tt hep-ph/0501277}}.

\bibitem{Son:1998uk}
D.~T. Son, {\it Superconductivity by long-range color magnetic interaction in
  high-density quark matter},  {\em Phys. Rev.} {\bf D59} (1999) 094019,
  [\href{http://xxx.lanl.gov/abs/hep-ph/9812287}{{\tt hep-ph/9812287}}].

\bibitem{Schafer:1999jg}
T.~Sch{\"a}fer and F.~Wilczek, {\it Superconductivity from perturbative
  one-gluon exchange in high density quark matter},  {\em Phys. Rev.} {\bf D60}
  (1999) 114033, [\href{http://xxx.lanl.gov/abs/hep-ph/9906512}{{\tt
  hep-ph/9906512}}].

\bibitem{Pisarski:1999bf}
R.~D. Pisarski and D.~H. Rischke, {\it Gaps and critical temperature for color
  superconductivity},  {\em Phys. Rev.} {\bf D61} (2000) 051501,
  [\href{http://xxx.lanl.gov/abs/http://arXiv.org/abs/nucl-th/9907041}{{\tt
  http://arXiv.org/abs/nucl-th/9907041}}].

\bibitem{Pisarski:1999tv}
R.~D. Pisarski and D.~H. Rischke, {\it Color superconductivity in weak
  coupling},  {\em Phys. Rev.} {\bf D61} (2000) 074017,
  [\href{http://xxx.lanl.gov/abs/nucl-th/9910056}{{\tt nucl-th/9910056}}].

\bibitem{Brown:1999aq}
W.~E. Brown, J.~T. Liu, and H.-c. Ren, {\it On the perturbative nature of color
  superconductivity},  {\em Phys. Rev.} {\bf D61} (2000) 114012,
  [\href{http://xxx.lanl.gov/abs/http://arXiv.org/abs/hep-ph/9908248}{{\tt
  http://arXiv.org/abs/hep-ph/9908248}}].

\bibitem{Brown:2000eh}
W.~E. Brown, J.~T. Liu, and H.-c. Ren, {\it Non-{F}ermi liquid behavior, the
  {BRST} identity in the dense quark-gluon plasma and color superconductivity},
   {\em Phys. Rev.} {\bf D62} (2000) 054013,
  [\href{http://xxx.lanl.gov/abs/http://arXiv.org/abs/hep-ph/0003199}{{\tt
  http://arXiv.org/abs/hep-ph/0003199}}].

\bibitem{Wang:2001aq}
Q.~Wang and D.~H. Rischke, {\it How the quark self-energy affects the
  color-superconducting gap},  {\em Phys. Rev.} {\bf D65} (2002) 054005,
  [\href{http://xxx.lanl.gov/abs/http://arXiv.org/abs/nucl-th/0110016}{{\tt
  http://arXiv.org/abs/nucl-th/0110016}}].

\bibitem{Altherr:1992mf}
T.~Altherr and U.~Kraemmer, {\it Gauge field theory methods for ultradegenerate
  and ultrarelativistic plasmas},  {\em Astropart. Phys.} {\bf 1} (1992)
  133--158.

\bibitem{Manuel:1996td}
C.~Manuel, {\it Hard dense loops in a cold non-abelian plasma},  {\em Phys.
  Rev.} {\bf D53} (1996) 5866--5873,
  [\href{http://xxx.lanl.gov/abs/hep-ph/9512365}{{\tt hep-ph/9512365}}].

\bibitem{Mrowczynski:2000fp}
S.~Mr{\'o}wczy{\'n}ski and M.~H. Thoma, {\it Hard-loop approach to anisotropic
  systems},  {\em Phys. Rev.} {\bf D62} (2000) 036011,
  [\href{http://xxx.lanl.gov/abs/hep-ph/0001164}{{\tt hep-ph/0001164}}].

\bibitem{Romatschke:2003ms}
P.~Romatschke and M.~Strickland, {\it Collective modes of an anisotropic quark
  gluon plasma},  {\em Phys. Rev.} {\bf D68} (2003) 036004,
  [\href{http://xxx.lanl.gov/abs/hep-ph/0304092}{{\tt hep-ph/0304092}}].

\bibitem{Arnold:2003rq}
P.~Arnold, J.~Lenaghan, and G.~D. Moore, {\it {QCD} plasma instabilities and
  bottom-up thermalization},  {\em JHEP} {\bf 08} (2003) 002,
  [\href{http://xxx.lanl.gov/abs/hep-ph/0307325}{{\tt hep-ph/0307325}}].

\bibitem{Mrowczynski:2004kv}
S.~{Mr\'owczy\'nski}, A.~Rebhan, and M.~Strickland, {\it Hard-loop effective
  action for anisotropic plasmas},  {\em Phys. Rev.} {\bf D70} (2004) 025004,
  [\href{http://xxx.lanl.gov/abs/hep-ph/0403256}{{\tt hep-ph/0403256}}].

\bibitem{Frenkel:1990br}
J.~Frenkel and J.~C. Taylor, {\it High temperature limit of thermal {QCD}},
  {\em Nucl. Phys.} {\bf B334} (1990) 199.

\bibitem{Braaten:1990mz}
E.~Braaten and R.~D. Pisarski, {\it Soft amplitudes in hot gauge theories: A
  general analysis},  {\em Nucl. Phys.} {\bf B337} (1990) 569.

\bibitem{Taylor:1990ia}
J.~C. Taylor and S.~M.~H. Wong, {\it The effective action of hard thermal loops
  in {QCD}},  {\em Nucl. Phys.} {\bf B346} (1990) 115--128.

\bibitem{Braaten:1992gm}
E.~Braaten and R.~D. Pisarski, {\it Simple effective {L}agrangian for hard
  thermal loops},  {\em Phys. Rev.} {\bf D45} (1992) 1827--1830.

\bibitem{Nadkarni:1988fh}
S.~Nadkarni, {\it Dimensional reduction in finite temperature quantum
  chromodynamics. 2},  {\em Phys. Rev.} {\bf D38} (1988) 3287.

\bibitem{KorthalsAltes:1999cp}
C.~P. Korthals-Altes, R.~D. Pisarski, and A.~Sinkovics, {\it The potential for
  the phase of the {W}ilson line at nonzero quark density},  {\em Phys. Rev.}
  {\bf D61} (2000) 056007, [\href{http://xxx.lanl.gov/abs/hep-ph/9904305}{{\tt
  hep-ph/9904305}}].

\bibitem{Hart:2000ha}
A.~Hart, M.~Laine, and O.~Philipsen, {\it Static correlation lengths in {QCD}
  at high temperatures and finite densities},  {\em Nucl. Phys.} {\bf B586}
  (2000) 443--474, [\href{http://xxx.lanl.gov/abs/hep-ph/0004060}{{\tt
  hep-ph/0004060}}].

\bibitem{Bodeker:2001fs}
D.~B{\"o}deker and M.~Laine, {\it Finite baryon density effects on gauge field
  dynamics},  {\em JHEP} {\bf 0109} (2001) 029,
  [\href{http://xxx.lanl.gov/abs/hep-ph/0108034}{{\tt hep-ph/0108034}}].

\bibitem{Blaizot:2001vr}
J.~P. Blaizot, E.~Iancu, and A.~Rebhan, {\it Quark number susceptibilities from
  {HTL}-resummed thermodynamics},  {\em Phys. Lett.} {\bf B523} (2001)
  143--150, [\href{http://xxx.lanl.gov/abs/hep-ph/0110369}{{\tt
  hep-ph/0110369}}].

\bibitem{Gavai:2001ie}
R.~V. Gavai, S.~Gupta, and P.~Majumdar, {\it Susceptibilities and screening
  masses in two flavor {QCD}},  {\em Phys. Rev.} {\bf D65} (2002) 054506,
  [\href{http://xxx.lanl.gov/abs/http://arXiv.org/abs/hep-lat/0110032}{{\tt
  http://arXiv.org/abs/hep-lat/0110032}}].

\bibitem{Gavai:2002kq}
R.~V. Gavai and S.~Gupta, {\it The continuum limit of quark number
  susceptibilities},  {\em Phys. Rev.} {\bf D65} (2002) 094515,
  [\href{http://xxx.lanl.gov/abs/http://arXiv.org/abs/hep-lat/0202006}{{\tt
  http://arXiv.org/abs/hep-lat/0202006}}].

\bibitem{Bernard:2004je}
{\bf MILC} Collaboration, C.~Bernard {\em et.~al.}, {\it Qcd thermodynamics
  with three flavors of improved staggered quarks},  {\em Phys. Rev.} {\bf D71}
  (2005) 034504, [\href{http://xxx.lanl.gov/abs/hep-lat/0405029}{{\tt
  hep-lat/0405029}}].

\bibitem{Gavai:2004sd}
R.~V. Gavai and S.~Gupta, ``The critical end point of {QCD}.'' hep-lat/0412035.

\bibitem{Allton:2005gk}
C.~R. Allton {\em et.~al.}, {\it Thermodynamics of two flavor qcd to sixth
  order in quark chemical potential},  {\em Phys. Rev.} {\bf D71} (2005)
  054508, [\href{http://xxx.lanl.gov/abs/hep-lat/0501030}{{\tt
  hep-lat/0501030}}].

\bibitem{Vuorinen:2003fs}
A.~Vuorinen, {\it The pressure of {QCD} at finite temperatures and chemical
  potentials},  {\em Phys. Rev.} {\bf D68} (2003) 054017,
  [\href{http://xxx.lanl.gov/abs/hep-ph/0305183}{{\tt hep-ph/0305183}}].

\bibitem{Allton:2003vx}
C.~R. Allton, S.~Ejiri, S.~J. Hands, O.~Kaczmarek, F.~Karsch, E.~Laermann, and
  C.~Schmidt, {\it The equation of state for two flavor {QCD} at non-zero
  chemical potential},  {\em Phys. Rev.} {\bf D68} (2003) 014507,
  [\href{http://xxx.lanl.gov/abs/hep-lat/0305007}{{\tt hep-lat/0305007}}].

\bibitem{Karsch:2000ps}
F.~Karsch, E.~Laermann, and A.~Peikert, {\it The pressure in 2, 2+1 and 3
  flavour {QCD}},  {\em Phys. Lett.} {\bf B478} (2000) 447,
  [\href{http://xxx.lanl.gov/abs/hep-lat/0002003}{{\tt hep-lat/0002003}}].

\bibitem{Fodor:2002km}
Z.~Fodor, S.~D. Katz, and K.~K. Szab{\'o}, {\it The {QCD} equation of state at
  nonzero densities: Lattice result},  {\em Phys. Lett.} {\bf B568} (2003)
  73--77, [\href{http://xxx.lanl.gov/abs/hep-lat/0208078}{{\tt
  hep-lat/0208078}}].

\bibitem{Szabo:2003kg}
K.~K. Szab{\'o} and A.~I. Toth, {\it Quasiparticle description of the {QCD}
  plasma, comparison with lattice results at finite {T} and {$\mu$}},  {\em
  JHEP} {\bf 0306} (2003) 008,
  [\href{http://xxx.lanl.gov/abs/hep-ph/0302255}{{\tt hep-ph/0302255}}].

\bibitem{Rebhan:2003wn}
A.~Rebhan and P.~Romatschke, {\it {HTL} quasiparticle models of deconfined
  {QCD} at finite chemical potential},  {\em Phys. Rev.} {\bf D68} (2003)
  025022, [\href{http://xxx.lanl.gov/abs/hep-ph/0304294}{{\tt
  hep-ph/0304294}}].

\bibitem{Ipp:2003yz}
A.~Ipp, A.~Rebhan, and A.~Vuorinen, {\it Perturbative {QCD} at non-zero
  chemical potential: Comparison with the large-{$N_f$} limit and apparent
  convergence},  {\em Phys. Rev.} {\bf D69} (2004) 077901,
  [\href{http://xxx.lanl.gov/abs/hep-ph/0311200}{{\tt hep-ph/0311200}}].

\bibitem{Moore:2002md}
G.~D. Moore, {\it Pressure of hot {QCD} at large {$N_f$}},  {\em JHEP} {\bf
  0210} (2002) 055, [\href{http://xxx.lanl.gov/abs/hep-ph/0209190}{{\tt
  hep-ph/0209190}}].

\bibitem{Ipp:2003zr}
A.~Ipp, G.~D. Moore, and A.~Rebhan, {\it Comment on and erratum to `{P}ressure
  of hot {QCD} at large {$N_f$}'},  {\em JHEP} {\bf 0301} (2003) 037,
  [\href{http://xxx.lanl.gov/abs/hep-ph/0301057}{{\tt hep-ph/0301057}}].

\bibitem{Ipp:2003jy}
A.~Ipp and A.~Rebhan, {\it Thermodynamics of large-{$N_f$} {QCD} at finite
  chemical potential},  {\em JHEP} {\bf 0306} (2003) 032,
  [\href{http://xxx.lanl.gov/abs/hep-ph/0305030}{{\tt hep-ph/0305030}}].

\bibitem{Blaizot:2005fd}
J.-P. Blaizot, A.~Ipp, and A.~Rebhan, {\it Study of the gluon propagator in the
  large-{$N_f$} limit at finite temperature and chemical potential for weak and
  strong couplings},  \href{http://xxx.lanl.gov/abs/hep-ph/0508317}{{\tt
  hep-ph/0508317}}.

\bibitem{Blaizot:2005wr}
J.-P. Blaizot, A.~Ipp, A.~Rebhan, and U.~Reinosa, {\it Asymptotic thermal quark
  masses and the entropy of {QCD} in the large-{$N_f$} limit},  {\em Phys.
  Rev.} {\bf D72} (2005) 125005,
  [\href{http://xxx.lanl.gov/abs/hep-ph/0509052}{{\tt hep-ph/0509052}}].

\bibitem{IKRV}
A.~Ipp, K.~Kajantie, A.~Rebhan, and A.~Vuorinen. Paper in preparation.


\bibitem{Freedman:1977dm}
B.~A. Freedman and L.~D. McLerran, {\it Fermions and gauge vector mesons at
  finite temperature and density. 2. {T}he ground state energy of a
  relativistic electron gas},  {\em Phys. Rev.} {\bf D16} (1977) 1147.

\bibitem{Freedman:1977ub}
B.~A. Freedman and L.~D. McLerran, {\it Fermions and gauge vector mesons at
  finite temperature and density. 3. {T}he ground state energy of a
  relativistic quark gas},  {\em Phys. Rev.} {\bf D16} (1977) 1169.

\bibitem{Baluni:1978ms}
V.~Baluni, {\it Non-abelian gauge theories of {F}ermi systems:
  Quantum-chromodynamic theory of highly condensed matter},  {\em Phys. Rev.}
  {\bf D17} (1978) 2092--2121.

\bibitem{Kap:FTFT}
J.~I. Kapusta, {\em Finite-temperature field theory}.
\newblock Cambridge University Press, Cambridge, UK, 1989.


\bibitem{LL:V}
L.~D. Landau and E.~M. Lifshitz, {\em Statistical Physics}.
\newblock Pergamon Press, London, 1958.


\bibitem{Holstein:1973}
T.~Holstein, R.~E. Norton, and P.~Pincus, {\it {d}e {H}aas-van {A}lphen effect
  and the specific heat of an electron gas},  {\em Phys. Rev.} {\bf B8} (1973)
  2649--2656.


\bibitem{Reizer:1989}
M.~Y. Reizer, {\it Relativistic effects in the electron density of states,
  specific heat, and the electron spectrum of normal metals},  {\em Phys. Rev.}
  {\bf B40} (1989) 11571--11575.


\bibitem{Gan:1993}
J.~Gan and E.~Wong, {\it Non-{F}ermi-liquid behavior in quantum critical
  systems},  {\em Phys. Rev. Lett.} {\bf 71} (1993) 4226--4229.


\bibitem{Chakravarty:1995}
S.~Chakravarty, R.~E. Norton, and O.~F. Sylju{\aa}sen, {\it Transverse gauge
  interactions and the vanquished {F}ermi liquid},  {\em Phys. Rev. Lett.} {\bf
  74} (1995) 1423--1426.

\bibitem{Varma:1989}
C.~M. Varma, P.~B. Littlewood, S.~Schmitt-Rink, E.~Abrahams, and A.~E.
  Ruckenstein, {\it Phenomenology of the normal state of cu-o high-temperature
  superconductors},  {\em Phys. Rev. Lett.} {\bf 63} (1989) 1996--1999.


\bibitem{Polchinski:1992ed}
J.~Polchinski, ``Effective field theory and the {F}ermi surface.''
  hep-th/9210046.

\bibitem{Polchinski:1994ii}
J.~Polchinski, {\it Low-energy dynamics of the spinon gauge system},  {\em
  Nucl. Phys.} {\bf B422} (1994) 617--633.

\bibitem{Nayak:1994ng}
C.~Nayak and F.~Wilczek, {\it Renormalization group approach to low temperature
  properties of a nonfermi liquid metal},  {\em Nucl. Phys.} {\bf B430} (1994)
  534--562.

\bibitem{Rischke:2003mt}
D.~H. Rischke, {\it The quark-gluon plasma in equilibrium},  {\em Prog. Part.
  Nucl. Phys.} {\bf 52} (2004) 197--296,
  [\href{http://xxx.lanl.gov/abs/nucl-th/0305030}{{\tt nucl-th/0305030}}].

\bibitem{Boyanovsky:2000bc}
D.~Boyanovsky and H.~J. de~Vega, {\it Non-{F}ermi liquid aspects of cold and
  dense {QED} and {QCD}: Equilibrium and non-equilibrium},  {\em Phys. Rev.}
  {\bf D63} (2001) 034016, [\href{http://xxx.lanl.gov/abs/hep-ph/0009172}{{\tt
  hep-ph/0009172}}].

\bibitem{Boyanovsky:2000zj}
D.~Boyanovsky and H.~J. de~Vega, {\it The specific heat of normal, degenerate
  quark matter: Non-{F}ermi liquid corrections},  {\em Phys. Rev.} {\bf D63}
  (2001) 114028, [\href{http://xxx.lanl.gov/abs/hep-ph/0011354}{{\tt
  hep-ph/0011354}}].

\bibitem{Ipp:2003cj}
A.~Ipp, A.~Gerhold, and A.~Rebhan, {\it Anomalous specific heat in high-density
  {QED} and {QCD}},  {\em Phys. Rev.} {\bf D69} (2004) R011901,
  [\href{http://xxx.lanl.gov/abs/hep-ph/0309019}{{\tt hep-ph/0309019}}].

\bibitem{Gerhold:2004tb}
A.~Gerhold, A.~Ipp, and A.~Rebhan, {\it Non-{F}ermi-liquid specific heat of
  normal degenerate quark matter},  {\em Phys. Rev.} {\bf D70} (2004) 105015,
  [\href{http://xxx.lanl.gov/abs/hep-ph/0406087}{{\tt hep-ph/0406087}}].

\bibitem{Schafer:2004zf}
T.~Sch{\"a}fer and K.~Schwenzer, {\it Non-{F}ermi liquid effects in {QCD} at
  high density},  {\em Phys. Rev.} {\bf D70} (2004) 054007,
  [\href{http://xxx.lanl.gov/abs/hep-ph/0405053}{{\tt hep-ph/0405053}}].

\bibitem{Iwamoto:1980eb}
N.~Iwamoto, {\it Quark beta decay and the cooling of neutron stars},  {\em
  Phys. Rev. Lett.} {\bf 44} (1980) 1637--1640.

\bibitem{Carter:2000xf}
G.~W. Carter and S.~Reddy, {\it Neutrino propagation in color superconducting
  quark matter},  {\em Phys. Rev.} {\bf D62} (2000) 103002,
  [\href{http://xxx.lanl.gov/abs/hep-ph/0005228}{{\tt hep-ph/0005228}}].

\bibitem{Schafer:2004jp}
T.~Sch{\"a}fer and K.~Schwenzer, {\it Neutrino emission from ungapped quark
  matter},  {\em Phys. Rev.} {\bf D70} (2004) 114037,
  [\href{http://xxx.lanl.gov/abs/astro-ph/0410395}{{\tt astro-ph/0410395}}].

\bibitem{Gerhold:2005uu}
A.~Gerhold and A.~Rebhan, {\it Fermionic dispersion relations in
  ultradegenerate relativistic plasmas beyond leading logarithmic order},  {\em
  Phys. Rev.} {\bf D71} (2005) 085010,
  [\href{http://xxx.lanl.gov/abs/hep-ph/0501089}{{\tt hep-ph/0501089}}].

\end{thebibliography}

\providecommand{\href}[2]{#2}\begingroup\raggedright\endgroup

\end{document}